\documentstyle[psfig,conf_iap]{article}
\begin{document}
\heading{%
%
Constraining the dark energy with Ly-$\alpha$ forest
%
} 
\par\medskip\noindent
\author{%
Uro\v s Seljak, Rachel Mandelbaum, Patrick McDonald

}
\address{%
Department of Physics, Princeton University, Princeton NJ USA
}

\begin{abstract}

Statistical properties of gas absorption in high redshift quasars 
such as power spectrum and 
bispectrum allow one to determine the evolution of structure 
over the redshift range $2<z<4$. Sloan Digital Sky Survey (SDSS) 
will measure 
around 10,000 quasar spectra in this redshift range and will allow one to 
determine the growth factor with a few percent accuracy.
This allows one to extend the studies of dark energy to high 
redshift and 
determine the presence of dark energy if $\Omega_{\rm de}>0.1-0.2$ 
at $z>2$. In combination with low redshift studies one can place
useful limits on the time evolution of the equation of state. 

\end{abstract}
\def\bi#1{\hbox{\boldmath{$#1$}}}
\def\gsim{\raise2.90pt\hbox{$\scriptstyle
>$} \hspace{-6.4pt}
\lower.5pt\hbox{$\scriptscriptstyle
\sim$}\; }
\def\lsim{\raise2.90pt\hbox{$\scriptstyle
<$} \hspace{-6pt}\lower.5pt\hbox{$\scriptscriptstyle\sim$}\; }
\def\Msun{\ifmmode {\rm\,M_\odot} \else
        ${\rm\,M_\odot}$\fi}

\section{Introduction }
The study of the Ly-$\alpha$ forest has been revolutionized in recent years
by the high resolution measurements using Keck HIRES spectrograph 
and by the development of
theoretical understanding using hydrodynamical simulations and
analytical models. The picture that has emerged from these studies
is one where the neutral gas responsible for the absorption is in a relatively
low density, smooth enviroment, which implies a simple connection between the gas
and the underlying dark matter. The neutral fraction of the gas is determined
by the interplay between the recombination rate (which depends on the
temperature of the gas) and ionization caused by
ultraviolet photons, the so called UV background. Photoionization heating
and expansion cooling cause the gas density and temperature to be tightly
related, except where mild shocks heat up the gas. This
leads to a tight relation between the absorption
flux and the gas density. Finally, the gas density is
closely related to the dark matter density on large scales, while on small
scales the effects of thermal broadening and Jeans smoothing have to
be included.
In the simplest picture described here all of the physics ingedients are
known and can be modelled.
In practice, the nonlinear physics requires the use of
hydrodynamic simulations with sufficient dynamic range which are only
now becoming available.

In the past couple of years cosmological observations 
such as redshift-luminosity relation from supernovae and 
position of the acoustic peaks in cosmic microwave background (CMB)
combined with local matter density estimates have revealed the presence 
of another component in the universe, the so called dark energy. 
This component has negative pressure and the recent 
constraints indicate that $w=p/\rho \sim -1$. However, at the 
moment direct constraints are still allowing for a significant 
range in $w$, specially if it is allowed to vary in time. Many models 
have been proposed where $w$ either increases or decreases with 
redshift (some of these are discussed in these proceedings). 
Theoretical studies have been performed on how to extract this evolution 
using low redshift probes.
While statistical power for some of these 
planned or proposed experiments, such as SNAP using supernovae, 
LSST using weak lensing or optical, X-ray and 
Sunyaev-Zeldovich telescopes counting clusters, 
is truly impressive, the control of systematics is less well 
understood. For this reason it is important to have as many independent 
tests as possible.

One possibility that has not been discussed much so far is using 
Ly-$\alpha$ forest as a probe of dark energy.
Since direct observations of quasars from the ground restrict one to 
the optical wavelengths ($\lambda>3600\AA$) one can only observe
Ly-$\alpha$ forest for $z>2$. If $w=-1$ independent of the redshift 
then if dark energy to dark matter density ratio 
$\Omega_{\rm de}/\Omega_m=2$ today the ratio will be below 0.1
at $z>2$. In this case the growth factor will scale almost 
linearly with the scale 
factor just as in an Einstein-de Sitter universe and one cannot detect 
the presence of dark energy.
On the other hand, if $w>-1$ either today or at a somewhat higher redshift 
then the decline of the dark energy fraction relative to dark matter 
is slowed down and dark energy can be dynamically important even at 
$z>2$. In this case there will be deviations in the growth factor 
from the expected scaling that can be detected using the forest 
observations. 

So far the only tests of the dark energy proposed have been using either
the growth factor or the redshift luminosity distance. If one wishes
to extract $w(z)$ then both of these involve a double integral over
this quantity and degeneracies arise. A more direct and still in 
principle observable way is to measure the Hubble parameter 
$H(z)$, which is related to $w(z)$ through a single integral. 
One way to measure it is to have a characteristic feature
which is fixed in comoving coordinates and observed in redshift space. 
The relation between redshift space and comoving space is the 
Hubble parameter itself and so observing the feature as a function 
of redshift provides $H(z)$ directly. The problem of course is that 
there are no characteristic features imprinted in large scale structure, 
since the distribution of structure
in the universe is stochastic in nature. 
One must therefore look for a characteristic 
scale in correlations between structures. 
In principle such features could be 
provided by baryonic oscillations imprinted in the matter power spectrum, but 
in practice this is a weak effect limited to very large scales and 
so cannot be made very precise. One is thus left with the 
variations in the correlation function slope as a function of scale. 
It is well known that the power spectrum slope varies from $n\sim 1$ 
on large scales to $n \sim -3$ on small scales independent of redshift
in CDM models. Hence the scale at which the 
slope takes a specific value can be viewed as a standard ruler and 
can be traced with redshift. If this slope is measured in redshift space, 
as is the case for either galaxy clustering along the line of sight 
or Ly-$\alpha$ forest correlations, then one is measuring 
directly $H(z)$. To be able to do this one has to detect the curvature
in the slope over the dynamic range of observations. This is 
challenging, since the dynamic range is narrow and the error on the 
slope will be large. We will show below that such a detection should be
possible with the current sample, but is not expected to improve the 
constraints on $w(z)$ significantly.

\section{Error estimation}

The basic approach to the error estimation is a standard one. 
Arranging the statistics one is estimating (in our case power spectrum 
and bispectrum) 
into a vector $\bf x$ and the parameters one is interested in into vector 
$\bf y$ the Fisher matrix is given by
\begin{equation}\label{E:fishdef}
F_{kl} = \left(\frac{\mathrm{d}\mathbf{x}}{\mathrm{d}y_k}\right)^T \cdot \mathsf{C}^{-1} \cdot \left(\frac{\mathrm{d}\mathbf{x}}{\mathrm{d}y_l}\right)
,
\end{equation}
where $C_{ij}=\langle (x_i-\langle x_i\rangle) (x_j-\langle x_j\rangle) \rangle$.

One must thus
compute both the covariance matrix and the derivatives with respect to 
the parameters on the statistic than one is 
using. 
We have found that combining the bispectrum and the power spectrum
information
significantly improves the determination of the amplitude of fluctuations.
The reason for this is a degeneracy between the mean flux absorption and 
the linear amplitude. 
Changing the mean flux changes the power spectrum of 
the flux significantly, as does the variation in the linear amplitude.
The mean flux absorption is a free parameter, since 
it is governed by the amount of UV background that controls the 
fraction of neutral gas in ionizing equilibrium. 
While it can be determined directly using independent methods such as 
continuum fitting or principal component analysis of spectra, the 
precision of these methods is not sufficient to break the 
degeneracies entirely. However, at the 3-point function level gravity 
predicts a very specific pattern of correlations, which cannot be 
mimicked by the mean flux variation. A full analysis reveals that this 
can improve the precision of amplitude determination by a factor of 3.

To compute the covariance matrix of power spectrum and bispectrum we use 
hydro-PM simulations to simulate the forest, apply the analysis 
on many realizations of the simulations and compute the mean and  
covariance matrix of the statistics. We use power spectrum 
and bispectrum information for $10^{-3}s/km<k<2\times 10^{-2}s/km$, 
which is the range of interest for the SDSS data. We then vary by 
small amount 
the parameters of interest one at a time, rerun the simulation with the
same initial conditions to minimize the sampling variance and 
recompute the mean. This allows us to compute 
the derivatives $dx_i/dy_j$ of the 
statistic $x_i$ with respect to the parameter $y_j$. 
The parameters we vary are the amplitude of fluctuations as a
function of redshift (we use 9 bins between $2.2<z<3.8$), 
mean flux as a function of redshift, slope and curvature of 
the primoridal spectrum and temperature density relation 
parametrized as a power law with amplitude and slope as the 
parameters, both of which can vary with redshift. 

The assumed redshift distribution of the sample mimics the SDSS 
distribution in the current data. We have scaled the length of the 
forest to match the overall number of quasars to 3000, which is 
significantly less than 
the final sample of around 10,000 QSOs. Our results are thus
conservative and will be improved significantly in the future, 
assuming that systematic errors can be kept under control. 
We assume the spectra have signal to noise of 5 and resolution 
of 70km/s, which is typical of the SDSS spectra. Neither signal to noise
nor resolution are particularly critical in this application, since 
one is mostly interested in large scale correlations where other 
parameters such as the temperature of the gas or the density temperature relation 
do not play  a major role. 

\section{Dark energy constraints}

To study the dark energy parameters we project the Fisher matrix to 
fewer parameters, parametrizing the growth factor evolution in terms 
of $\Omega_{\rm de}$ and $w(z)$. For the redshift evolution we limit 
ourselves to constant and linear evolution models, $w_q = w_0 + w_1 (a-1)$. 
The density in dark energy and equation of state are degenerate if one 
only uses information from Ly-$\alpha$ forest. This is not surprising, 
since the redshift range probed is too small to determine two parameters
from the growth factor evolution (as mentioned above, we find that 
$H(z)$ information does not provide any additional constraints). 
In the following we fix  
$\Omega_{\rm de}$ and present errors on equation of state only.  
The motivation for this is that
other tests, most notably CMB combined with large scale structure tests
(e.g. cluster counts, galaxy clustering, weak lensing), 
will be able to determine $\Omega_{\rm de}$ at $z=0$ very accurately.
If these tests also provide independent constraints on $w_0$
then for the time dependent $w$ one can use our results to constrain $w_1$. 

Since the sensitivity to the dark energy depends strongly on the amount 
of dark energy at $z>2$ it is clear that the errors will depend strongly 
on the assumed values of $\Omega_{\rm de}$ and $w(z)$. For the models studied
here their values are given in 
table \ref{T:qmodels}. 
The table also shows the errors on $w_0$ assuming fixed
$\Omega_{\rm de}$ (and $w_1$ in time dependent models) and errors
on $w_1$ assuming fixed $\Omega_{\rm de}$ and $w_0$. The errors 
are marginalized over all the other parameters.
One can see that the limits improve if $w_0$ is more positive or if
$w_1$ is significantly negative, since the dark energy is then 
more important at higher redshift. 
Of particular interest are
the limits on time dependent $w$. For example, in models 7 and 8 we assume 
today $w_0=-0.8$, which increases to $w=-0.4$ and $w=-0.2$ at z=2.6, 
respectively.  
In such models the error on $w_1$ is 0.2, which makes them 
distinguishable at 1.5$\sigma$ with the current sample and 3$\sigma$ with 
the full sample, assuming that both $\Omega_{\rm de}$ and $w_0$ can 
be accurately determined with the other methods. These errors improve further
if we live in a universe with lower matter density than 
$\Omega_m=0.33$ assumed here. 

While there is no 
simple single parameter combination that describes the sensitivity 
to dark energy it is clear that the precision 
is correlated 
with $\Omega_{\rm de}$ at $z=2.6$. Our results show that if 
$\Omega_{\rm de}(z=2.6)>0.2$ then the deviations in the growth 
factor are sufficiently large to be detected in Ly-$\alpha$ forest 
spectra using the current SDSS sample. With the full SDSS sample 
this limit can be improved further and models with $\Omega_{\rm de}(z=2.6)>0.1$
should be detectable. While this is still not sufficiently accurate 
to measure dark energy directly for cosmological constant model ($w=-1$) 
$\Omega_m \sim 0.3$, it will provide important constraints on the 
more general models of dark energy such as the tracker models, where 
equation of state naturally increases in value at higher redshifts. 

\begin{table}
\begin{center}
\begin{tabular}{| c | c | c | c |  c | c | c |}
Model & $\Omega_{q,0}$ & $\Omega_q(z=2.6)$ & $w_0$ & $w_1$ &  $\sigma_{w_0}$ & $\sigma_{w_1}$\\ \hline
1 & 0.67 & 0.12 & -0.7 & 0.0 &  0.22 & -\\
2 & 0.85 & 0.28 & -0.7 & 0.0 &  0.09 & -\\
3 & 0.49 & 0.06 & -0.7 & 0.0 & 0.36 & -\\
4 & 0.67 & 0.04 & -1.0 & 0.0 & 0.36 & -\\
5 & 0.67 & 0.30 & -0.4 & 0.0 &  0.17 & - \\
6 & 0.67 & 0.12 & -0.8 & -0.2 & 0.12 & 0.31\\
7 & 0.67 & 0.19 & -0.8 & -0.5 & 0.09 & 0.22 \\
8 & 0.67 & 0.27 & -0.8 & -0.8 & 0.068 & 0.18\\
9 & 0.67 & 0.09 & -1.0 & -0.5 & 0.16 & 0.41\\
\end{tabular}
\caption{Parameters and error bounds 
for dark energy models, where $w_q=w_0+w_1 (a-1)$.The errors are marginalized
over all the other parameters except dark energy ones (see text for details). }
\label{T:qmodels}
\end{center}
\end{table}


%
\vfill
\end{document}